\documentclass[12pt]{article}
\usepackage[utf8]{inputenc}
\usepackage{authblk}
\usepackage{natbib}
\usepackage{graphicx}
\usepackage{adjustbox}
\usepackage{geometry}
\usepackage{subcaption}
\usepackage{hyperref}

\title{\Large{\textbf{Biblical names' relationships in the Gospel of Matthew, Mark, Luke, John and Acts of Apostles\footnote{We are thankful to Prof. Marco V. Fabbri and Valentina Ferrigno for their effective suggestions.}}}}

\author[1]{\large{Roberto Rondinelli}}
\author[2]{\large{Stefano Marmani}}
\author[2]{\large{Valerio Ficcadenti}}

\affil[1]{\normalsize{University of Macerata - r.rondinelli@unimc.it}}
\affil[2]{\normalsize{London South Bank University - marmanis@lsbu.ac.uk; ficcadv2@lsbu.ac.uk}}
\date{\normalsize{April 2020}}

\begin{document}

\maketitle

\begin{center}
\section*{\normalsize{Abstract}}  
\end{center}
\footnotesize{In this paper we extrapolate the information about Bible’s characters and places, and their interrelationships, by using text mining network-based approach. We study the narrative structure of the WEB version of 5 books: the Gospel of Matthew, Mark, Luke, John and Acts of the Apostles. The main focus is the protagonists' names interrelationships in an analytical way, namely using various network-based methods and descriptors. This corpus is processed for creating a network: we download the names of people and places from Wikipedia's list of biblical names, then we look for their co-occurrences in each verse and, at the end of this process, we get $N$ co-occurred names. The strength of the link between two names is defined as the sum of the times that these occur together in all the verses, in this way we obtain 5 adjacency matrices (one per book) of $N$ by $N$ couples of names. 
After this pre-processing phase, for each of the 5 analysed books we calculate the main network centrality measures (classical degree, weighted degree, betweenness and closeness), the network vulnerability and we run the Community Detection algorithm to highlight the role of Messiah inside the overall networks and his groups (communities). We have found that the proposed approach is suitable for highlighting the structures of the names co-occurrences. The found frameworks' structures are useful for interpreting the characters' plots under a structural point of view.}

\vspace{4mm}
\normalsize{\textbf{Keywords:}} \footnotesize{Characters Network, Text Mining, Acts of Apostles, Gospels}
\vspace{3mm}

\section*{\large{1. Introduction}}

There are many studies in the literature which address texts’ frameworks via word-based networks \citep[see for example][where such an approach has been computerized for one of the firsts times]{knuth1993stanford}.
Sometimes words are considered as nodes and their co-occurrences in the texts as their links. In other contexts, texts are intended as nodes and the number of shared words as their links. In the same way, the authors can constitute the nodes of the network when corpora are studied.\\
In this paper we have used a text mining network-based approach to study the relationship between proper names in the Gospel of Matthew, Mark, Luke, John and Acts of the Apostles from the World English Bible\footnote{World English Bible is a free updated revision of the American Standard Version of bible which started on 1901} \citep[WEB hereafter, see][]{WEB}, with a specific focus on the Messiah's names and their neighbourhood framework.
We have decided to use such a version of the Bible because a recent translation of the original text can help embedding all the latest findings on names' translations. Indeed, most of the biblical proper names carry out meanings, therefore different translations of them can be met (see \citealt{vanhoozer2005dictionary} which is a report of facts' interpretations, or, for example, see the different meanings for \textit{``Abba"} in \citealt{smith1967smith} and \citealt{orr1939international} ).
The idea of studying the characters' relationships via a text mining network-based approach is used to verify some well-known positions of the figures in the biblical history, with a specific focus on the Messiah’s centrality.  We investigate the difference between the Acts of Apostles structures and the other Gospels and the references to Old Testament episodes in the books.\\
The study contains Section 2 for the related works and Section 3 where a description of the data collection process and the formation of the adjacency matrices is reported. Section 4 reports the network's analytic tools used, in Section 5 are presented results and comments. Finally, in Section 6, the conclusions are summarized.

\section*{\large{2. Related Studies}}

The study of the Bible is a meeting point for many disciplines, therefore it is the perfect ground for interdisciplinary research activities. It is worth to mention the noticeable cases of \cite{Busa1974BUSITS,bamman2008annotation,passarotti2007verso}, where the authors have contributed to make analysable by machines the works of Thomas Aquinas\footnote{see \url{https://www.corpusthomisticum.org/wintroen.html} and \url{https://itreebank.marginalia.it}}.
In the case of this study, we are addressing the narrative paths of the aforementioned 5 books by means of the names' presences. The tools for the analysis come from the field of text mining ( IT / data science) and the analytic are borrowed from the network theory (mathematical graph theory).
This approach is getting attention because of the potentialities it offers. In \cite{cinelli2019interconnectedness} there is an example of information gained by the study of US President speeches using a similar approach. In \cite{Rochat2014} the authors have studied the central figures of Les Confessions (Geneva version of the manuscript) and in \cite{Rochat2015} has been presented an excellent study of the characters' framework in the twenty novels composing Les Rougon-Macquart, written by \'Emile Zola between 1871-1893.

Referring to biblical studies, from a network theory point of view, we are in line with \cite{czachesz2016network,zhitomirsky2019sagebook} even but we add a Community Detection analysis to the common centrality measures. From a text mining point of view, we differ from \cite{czachesz2016network} because we do not build a semantic network as the author did, therefore we did not deal with stemming and lemmatization. Going deeper in the differences between the two approaches, instead of using almost all the words (semantic approach), we have used the list of names from Wikipedia's web page (\citealt{wiki:listofbiblicalnames}) which contains a wide catalogue of names mentioned in the Bible. Furthermore, this list have been checked against the proper names of the $28th$ tagged edition of the Nestle-Aland's Novum Testamentum Graece (see \citealp{Forte2013} for some observation and a description of this version) and versus the names listed by Christoph Romhild who recorded 1773 proper nouns – people and places – in the King James Bible, see \cite{namesdata1,namesdata2}.
Similarly to \cite{czachesz2016network}, we had to deal with the stop words.  The Wikipedia's list of biblical names have been assessed manually to remove potential problem like names which belong to the most common stop words lists.  For example ``So" could not be easily distinguished by the conjunction/adverb ``so", therefore it has been removed from the searching list (see \citealt{wiki:listofnameswithS} for the names starting with ``S" and \citealt{smith1967smith} for a definition of the word). In fact, for each verse of the 5 books of the New Testament (WEB version), we have looked for the names we found on the Wikipedia's list; once we have found them, we counted their co-occurrences in the same verses. In this way, we could create a network\footnote{For example, in Matthew 4:12 one finds \textit{``Now when Jesus heard that John was delivered up, he withdrew into Galilee"}, hence there is a link between Jesus and John, one between Jesus and Galilee and the last one between John and Galilee.}.
This said, our study is still comparable to Czachesz's for the shared approach.

If one considers the organization of the Bible and the books studied here, the smallest unit of analysis to be addressed, from a structural study point of view, is the verse. In making a decision we have followed the suggestions of \cite{Efrat1980} which reports \textit{``Structure can be defined as the network of relations among the parts of an object or a unit. This definition at once raises the question what is to be considered a unit in the area of biblical narrative"}. In his structural analysis, Bar-Efrat often refers to occurrences in verses, in particular when he investigates the characters' relationships.

Besides, we are in line with \cite{beveridge2016network}, whose study is very close to ours, both for the text mining and network analysis, but performed on the Game of Throne's books. In \cite{beveridge2016network}, the authors have decided to divide the text up into chunks using 15 word-long pieces. We are aware that both the methods have drawbacks due to loss of information connected to the fact that conceptual forms are not necessarily close in a single verse, sentence or chunk. In fact, we do not claim to perform a fully automated structural analysis in this paper, but we want to explore the capacity of this text mining network-based approach in comparison to more labour-intensive investigation.

Another relevant work in the field is \cite{massey2016social} where a \textit{``narrative network"} of the Pentateuch has been built to study the role of Moses. The author has made the network on the basis of the character \textit{``collaborations"}, namely \textit{``addressing directly or indirectly another character, family connection (parent-offspring and marriage, as with genealogical networks), inheritance, physical conflict or criticism"}. From a network analysis point of view, the two studies are comparable because both are interested in the role of a Messiah, even though \cite{massey2016social} focused more on studying the resulting networks via scale-free analytical tools, like in the case of the power laws fit of the nodes’ degree (preferential attachment mechanism). Furthermore, Massey investigates the network properties via assortativity measurement. In this stage of our study, we did not deepen the research until such a point because we prefer to devote attention to the networks' basic features and community structures.

\section*{\large{3. Data}}
\label{Data}

In this work the version of bible used is the World English Bible \citep{WEB}. It has been downloaded from \cite{ wiki:webbiblesql}\footnote{\url{https://www.kaggle.com/oswinrh/bible}}. The website distributes the 66 versions of the books (Old and New Testament) of the bible already organized in verses in SQL format. The object of our analysis is focused on the 4 Gospels (Matthew, Mark, Luke and John) and the Acts of Apostles because they are the most popular in European countries.
We are interested in studying the names which occurred in such books; therefore, we have looked for a list of biblical names to search in them. On Wikipedia \citep{ wiki:listofbiblicalnames} there is a complete and wide list of names. The pages associated with the main one\footnote{https://en.wikipedia.org/wiki/List\_of\_biblical\_names} contain 3094 words not only made of personal names but also names of places, names of spiritual and mythological entities. All of them significant for the Bible. 
We have assessed such list by searching the meaning of all the terms, one by one, and we have identified 3 names which can not be distinguished from other English words. They are \textit{``so", ``on", ``no"} and belong to the most common Stop Words lists. Keeping them in the list could bias the results due to their high frequency of occurrence. In fact, their presence is so wide that their co-occurrences pair with almost all the characters, therefore their connections can be considered as noise in our analysis.
Each name has been searched in all the verses to register the co-occurrence between them. Indeed, to build the network, we have registered the names occurring together in the verses and we have marked them as linked. The names are the nodes of the network and their connections are weighted via the number of times couples appear in the verses. For example, from Matthew 12:1-3 \textit{``Then six days before the Passover, Jesus came to Bethany, where Lazarus was, who had been dead, whom he raised from the dead. 2 So they made him a supper there. Martha served, but Lazarus was one of those who sat at the table with him. 3 Therefore Mary took a pound of ointment of pure nard, very precious, and anointed Jesus’s feet and wiped his feet with her hair. The house was filled with the fragrance of the ointment."} the results are reported in Table \ref{example}.

    \begin{table}[ht]
\centering
\begin{tabular}{|c|c|c|c|c|c|}
\hline
        & Jesus & Bethany & Lazarus & Martha & Mary \\
        \hline
Jesus   &  -      & 1     & 1     &  -     &    1    \\
\hline
Bethany   & 1       & -     & 1     & -     &   -   \\
\hline
Lazarus   & 1       & 1     & -     & 1     &   -   \\
\hline
Martha   & -       & -     & 1    & -    & -     \\
\hline
Mary   &    1    &  -     &  -     & -     & -       \\
\hline
\end{tabular}
\caption{\small{\textit{The results from the network building process applied to Matthew 12:1-3}}}
\label{example}
\end{table}

With this procedure, one finds out that most names present in the original list do not occur together in the same verses in the books here considered. Therefore, they are not part of the resulting adjacency matrix. Such a table is made by 322 rows (in the 5 books together), each one containing a name occurred at least once with another name of the list. The table's content is the frequency of co-occurrence in the verses, namely the number of verses in which the couples have appeared. The resulting matrices are made by $N$ x $N$ elements, where $N$ corresponds to the number of nodes reported in Table \ref{netstats}. From the same table, it is possible to notice high sparsity (opposite of density) of the networks which foster the analysis. High sparsity means having the number of connections between nodes reasonably low, namely eligible to carry information about the characters' relationships in the analysed story. A fully sparse network would have nodes without connections while a complete network would have of all the nodes connected between each other. Both these situations are not ideal to perform text mining network-based analysis. In this study we are interested in observing the meaningful connections between the nodes of the network, therefore zero connections or maximum connectivity would not provide hints about the story's structure.\\

\begin{table}[htbp]
  \centering
    \begin{tabular}{|l|c|c|c|c|c|c|}
    \hline
  \multicolumn{1}{|r|}{} & \textbf{5 books} & \textbf{Acts} & \textbf{John} & \textbf{Luke} & \textbf{Mark} & \textbf{Matthew} \\
  \hline
    \textbf{Density} & 0.021 & 0.035 & 0.09  & 0.039 & 0.114 & 0.047 \\
    \hline
    \textbf{\# Nodes} & 322   & 176   & 58    & 128   & 45    & 104 \\
    \hline
    \end{tabular}%
  \caption{\small{\textit{Main indicators of the networks structures}}}
  \label{netstats}%
\end{table}%

\section*{\large{4. Network Analysis methods}}
\label{NetworkAnalysismethods}

For the analysis of the networks’ structure obtained by the textual pre-processing phase previously described, we make use of different network analytic tools: degree centrality, betweenness centrality, closeness centrality, network vulnerability and community detection.\\
Degree centrality requires  the number of links held by each node. The higher the node’s degree is (a lot of direct links with the other nodes), the more central (popular) it is in the network. Names or places (vertices in the network) with high degree are highly representative of the history told, therefore they are often recalled. Having many links with many other nodes means co-occurring in the same verse and therefore having relevant connections for the narrative.  In contrast, low-grade actors and places represent peripheral positions in the network, namely more marginal episodes, not always taken as reference in the story reported in the books.
We use both standard degree centrality (counting the number of neighbours) and the weighted degree centrality (based on the weights of the links).

Betweenness centrality measures the number of times a node lies on the shortest path between the other nodes. Nodes with high betweenness are considered to be mediators and they can influence and control the flow of information around the network. High level of betweenness could indicate that a node holds authority, controls collaboration between disparate groups or is in the periphery of both groups. \\
Closeness centrality indicates the ``closeness" of each node to all the other nodes of the network. In terms of geodesic distance (the shortest path of links between two nodes), the closer a node is to many others the more central it is in the network (it can quickly interact with the other actors). This centrality measure is particularly useful in more sparse graphs like these. \\ 
For further details on Degree, Betweenness and Closeness centrality see \cite{wasserman1994k}.\\
Network vulnerability (resilience and robustness of graphs) indicates the strength of a network maintaining its structure when nodes or edges are deleted from it. Here, we measure it in different ways (according to the NetSwan Rpackage, see \citealt{lhomme2015analyse}):

\begin{itemize}
    \item Loss in connectivity: the change in the sum of geodesic distances between all node pairs, deleting each node (see \citealt{lhomme2015analyse}).
    \item Loss in closeness: the change in the sum of the inverse of distances between all node pairs, deleting each node (see \citealt{lhomme2015analyse}).
    \item Loss in connectivity: the proportion of change in the sum of the inverse of distances between all node pairs, deleting nodes in decreasing order of their betweenness (see \citealt{albert2000error}).
    \item Loss in connectivity: the proportion of change in the sum of the inverse of distances between all node pairs, deleting nodes in decreasing order of their degree (see \citealt{albert2000error}).
    \item Loss in connectivity: the proportion of change in the sum of the inverse of distances between all node pairs, deleting nodes according to a cascading scenario where betweenness is recalculated after each node removed (see \citealt{albert2000error}).
    \item Loss in connectivity: the proportion of change in the sum of the inverse of distances between all node pairs, deleting nodes at random (see \citealt{albert2000error}).
\end{itemize}

To perform the community detection we employed the Louvain method (see \citealt{blondel2008fast}). It is a heuristic method based on modularity optimization. The way in which the algorithm works is typical in the cluster analysis method. Namely, it looks for the best partition maximizing the value of modularity.
This compares the network with a random graph in which the degree structure is kept the same as the original network (edges rewiring). Its theoretical value ranges in the interval [-1,1] and the higher it is the more evident the community structure is. 
The reasons for choosing this algorithm ground on its performances in terms of speed and on its ability of reaching better-defined communities with respect to other algorithms.

\section*{\large{5. Results and discussions}}
\label{results}

\begin{figure}
\centering
\includegraphics[width=\textwidth]{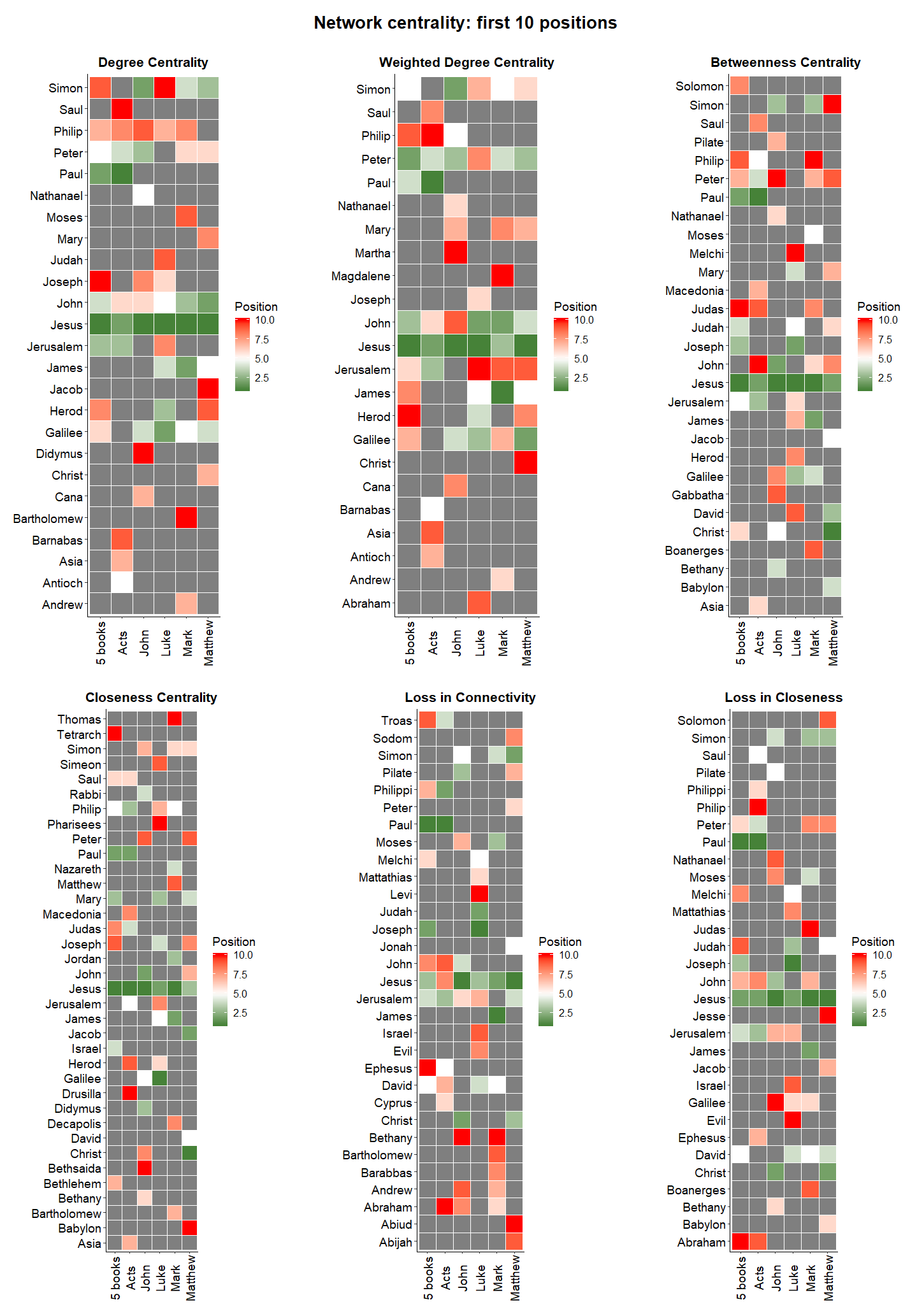}
\caption{\small{\textit{Comparing the first 10 names (nodes) with the highest centrality value measured via Degree, Weighted Degree, Betweenness, Closeness, loss in Connectivity and loss in Closeness. The color grey indicates that the name is not in the first 10 positions in that book according to the centrality measure.}}}
\label{heatmap}
\end{figure}

\begin{figure}
\centering
\includegraphics[width=\textwidth]{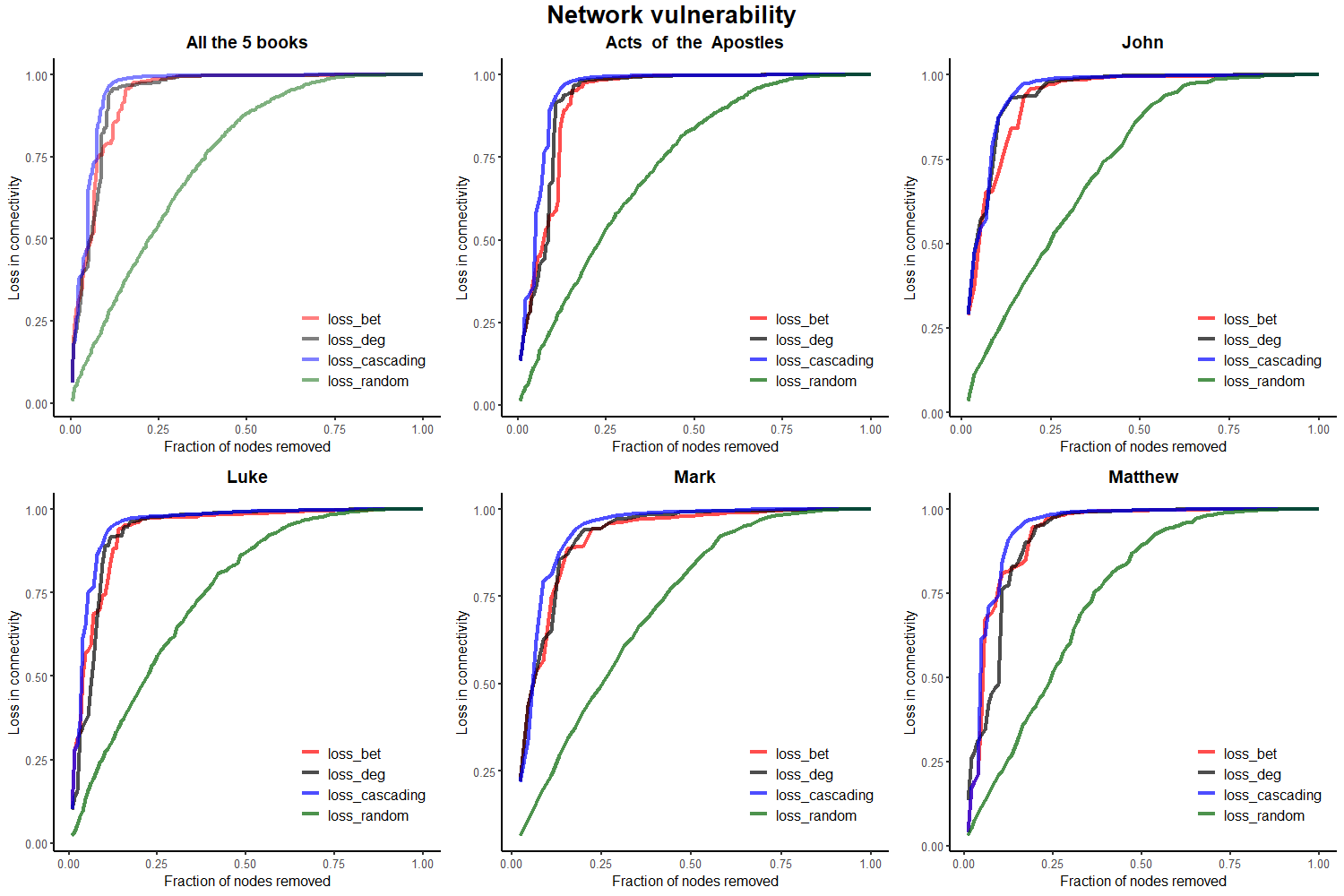}
\caption{\small{\textit{Evaluating the shape of network loss in Connectivity deleting nodes: from the highest value of Betweenness, the highest value of Degree, the highest value of Betweenness recalculating it for each removal nodes (cascading scenario), at random.}}}
\label{vulnerability}
\end{figure}

The results of the centrality measures are reported in Figure \ref{heatmap}. The colours represent the rank of the names when the list is sorted according to the centrality measure used. If on the rows referring to a certain name, one finds the same green shaded cells in the different books, it means that such a name is central in each text. When the colour of a certain name is the same across the different measurements (sub-plot of Figure \ref{heatmap}), it means that such a name has different centrality's features.  One of the most remarkable cases is obviously the Messiah, in fact looking at \textit{``Jesus"} we have remarkable relevance in all the books for the Degree, Weighted Degree, Betweenness and Closeness Centrality. Namely, the writers of the books have mentioned the Messiah in such a way that he has frequent connections with almost all the other names. This fact makes Jesus the character who connects people and places justifying also the results of Betweenness and Closeness Centrality. Loss in connectivity and loss in closeness highlight different aspects due to the nature of the measures.
The former confirms the Jesus' network importance for Gospel of John, Luke, Mark and Matthew, but, reasonably, in the Act of Apostles, the Messiah changes his position given the structure of the narrative. The Gospel written by Luke describes the rapid development, expansion and organization of the Christian testimony first to the Jews and then to the men of each nation. In this context, the name \textit{``Paul"} is fundamental for maintaining the level of connections between the names in the Act of Apostles (see high/green level of the name in Loss in Connectivity and so in the others measures; see also \citealt{greenwood1995common} where similar results are presented). Comparing Paul and Jesus in the particular cases of Act of Apostles, the relevance of the former is even more crucial than the latter's. This idea is well established in the literature, see for example \cite{TheJesusPaulParallelsand1975, hengel2003between}. These findings are visible in Figure \ref{networks} as well. The Act of Apostles clearly presents a second core of connections on the top right hand corner, namely around Paul. This and others nodes with their connections make the network more dense than the others.\\
The results of the vulnerability analysis are showed in Figure \ref{vulnerability}. The vulnerability analysis can be considered as a way of understanding how important the most central names are in the networks. The first element to be noticed is given by the changes in the starting points of the lines in each sub-figures. In fact, the x and y axes give an idea of the framework’s portion that the networks lose each time a central name is excluded.

\newgeometry{top=15mm}

\restoregeometry

Therefore, the highest the first line appears the more important Jesus is in that network (as we described above, an exception is constituted by the Act of Apostles because of Paul's role). From a structural point of view, these graphs prove that a core of nodes is crucial for maintaining the level of density's connections. Such an element might be a hint of the core-periphery framework's presence.

Another object of our analysis is the investigation of the communities of names that are present in the texts, with a specific focus on the Messiah. In Figure \ref{networks} and in Table \ref{community} a summary of the Louvain's detection algorithm outcomes is showed. In the table, per each book the modularity, the number of nodes and the resulting clusters are reported. The modularity goes between -1 and 1, the lower bound means poor network's ability to be organized in clusters, while 1 means the contrary. As it is possible to see, the modularities are quite high, strengthening the idea of having well-defined clusters.\\
Focusing on the clusters to which the Messiah belongs (see the highlighted elements in Figure \ref{networks}), we have noticed some relevant points. Table \ref{jesusclusters} is helpful for commenting on them. Firstly, one can notice \textit{``Jesus Christ"} does not appear with other names in Luke and Mark's Gospels, furthermore, it never appears with Jesus in the same verse, proving that they are the same character. While, when Christ is used without Jesus, it is used to testify Jesus' role. In Mark, Jesus and Christ appear in the same community together with most of the characters' names appeared to the last part of the Messiah's life (see Figure \ref{networks}). While in Luke, Jesus is preponderant in the cluster 12, there is no trace of Christ in it but the names' of some Apostles appear. Secondly, when the 5 books are considered together, the three names Jesus, Christ and Jesus Christ fall in the same cluster (the number 4 in brackets represent the cluster's number in Table \ref{jesusclusters}). Strangely, in this community, there are not apostles' names, but there other important names associated to \textit{``Barabbas"}, \textit{``Bethlehem"} and \textit{``Martha"} (these are just a few, given that the cluster number 4 is made of 61 names). Thirdly, in the Acts of Apostles the names Jesus and Christ belong to the community 2, where the remarkable names appearing are \textit{``Egypt"}, \textit{``Pharaoh"}, \textit{``Moses"} while the Apostles' names are missing. This is mostly given by the need of centralizing the role of Jesus as well as his salient actions.
In the Gospel of John, the names considered in Table \ref{jesusclusters} fall in different clusters. Jesus occurs with noticeable places like \textit{``Galilee", ``Israel"} and \textit{``Jerusalem"}. Differently, Christ belongs to a small group made of \textit{``Bethlehem"}, \textit{``David"} and \textit{``Elijah"}. The case of Matthew introduces another variation because Christ and Jesus Christ occur together in cluster mostly made by names connected with the Old Testament (e.g. \textit{``Eleazar"} or \textit{``Zerubbabel"}).

\begin{table}[htbp]
  \centering
    \begin{tabular}{|l|c|c|c|c|c|c|}
    \hline
   \multicolumn{1}{|r|}{} & \textbf{5 books} & \textbf{Acts} & \textbf{John} & \textbf{Luke} & \textbf{Mark} & \textbf{Matthew} \\
   \hline
    \textbf{modularity} & 0.49  & 0.51  & 0.32  & 0.64  & 0.43  & 0.56 \\
    \hline
    \textbf{\# nodes} & 322   & 176   & 58    & 128   & 45    & 104 \\
    \hline
    \textbf{\# clusters} & 23    & 12    & 7     & 16    & 6     & 9 \\
    \hline
    \end{tabular}%
  \caption{\small{\textit{Summary of the Community Detection results. Modularity value, number of nodes (names) and number of generated clusters.}}}
  \label{community}%
\end{table}%

\begin{table}[htbp]
  \centering
    \begin{adjustbox}{max width=\textwidth}
    \begin{tabular}{|l|c|c|c|c|c|c|}
    \hline
   \multicolumn{1}{|c|}{\textbf{Names}} & \textbf{5 books} & \textbf{Acts} & \textbf{John} & \textbf{Luke} & \textbf{Mark} & \textbf{Matthew} \\
   \hline
    \textbf{Jesus} & 0.19 (4) & 0.15 (2) & 0.33 (3) & 0.16 (12) & 0.35 (4) & 0.26 (4) \\
    \hline
    \textbf{Christ} & 0.19 (4) & 0.15 (2) & 0.07 (1) & 0.09 (8) & 0.35 (4) & 0.17 (5) \\
    \hline
    \textbf{Jesus Christ} & 0.19 (4) & 0.17 (5) & 0.17 (6) & 0     & 0     & 0.17 (5) \\
    \hline
    \end{tabular}%
  \end{adjustbox}
      \caption{\small{\textit{In each cell there is the weight of the cluster in which the name is contained and its membership cluster.}}}
  \label{jesusclusters}%
\end{table}%

\section*{\large{6. Conclusions}}
\label{conclusion}
The present study consists of an alternative approach for addressing the structural analysis of corpus which contains stories. With the help of network metrics usually applied in social network analysis and employing some text mining tools, we were able to identify key aspects of the Bible's names distributions and occurrences. Inferring from the results about the characters' plots into the targeted history, one could see overlap with the result already reported by scholars which have used classical structural analysis \citep[e.g.][]{barthes1974structural}. Enlarging this concept, an approach like the one proposed by Vladimir Propp in Morphology of the Folktale \citep{propp2010morphology} requires a combination of techniques which are close to those here presented. In fact, the narrative structures analysis is particularly  systematic, therefore the text mining network-based approach shows potentialities. Furthermore, the idea of having names of places as well as names of people in the analysis might allow the identification of structure like those presented in \cite{campbell2008hero}, for studying characters paths (e.g. ``The Hero's Journey").

In the previous section we can observe the Messiah's different positions in the Acts of Apostles with respect to the 4 Gospels (see for example \citealt{roetzel1999paul} where similar comments are presented). This is an expected result given the different natures and scopes of the books. The idea of highlighting it with tools designed to investigate structures (networks' analytical tools, see Section 4), emphasizes the fundamentals of structuralism making the plot of the story central and identifying it under a positivist and analytic light. This is particularly evident if one thinks about the design of centrality measures used and so their outcomes. \\
A similar logic applies to the results coming from vulnerability analysis and community detection. The former presents the importance of the most central names and it gives an idea of the relevance of the  connections in the network. The latter provides insights about the most interacting names within books. The results presented above show that sometimes the communities come from specific episodes like recalls from the Old Testament or Messiah and apostles interactions.

This type of approach presents findings which are well-known in the literature and it cannot substitute the work done by experts in the field. Rather, the network analysis approach manifests potentialities in supporting the work of such researchers, mitigating the necessary efforts to perform a systematic analysis of the corpus.

The Bible is a complex book adjusted along with the history and widely commented by scholars. Some fundamental facts find a general agreement in the literature and never drastically changed. This research has met some of these findings, but further research might be designed from here for a more systematic assessment of biblical events. One could test the results against a modern Greek version of the Bible, which should have higher adherence to the original text. This could improve the agreement about the names here selected, improving the precision of the adjacency matrix.

\bigskip
\bigskip

\begin{figure}[!hb]

  \begin{subfigure}[b]{0.5\textwidth}
    \includegraphics[width=\linewidth]{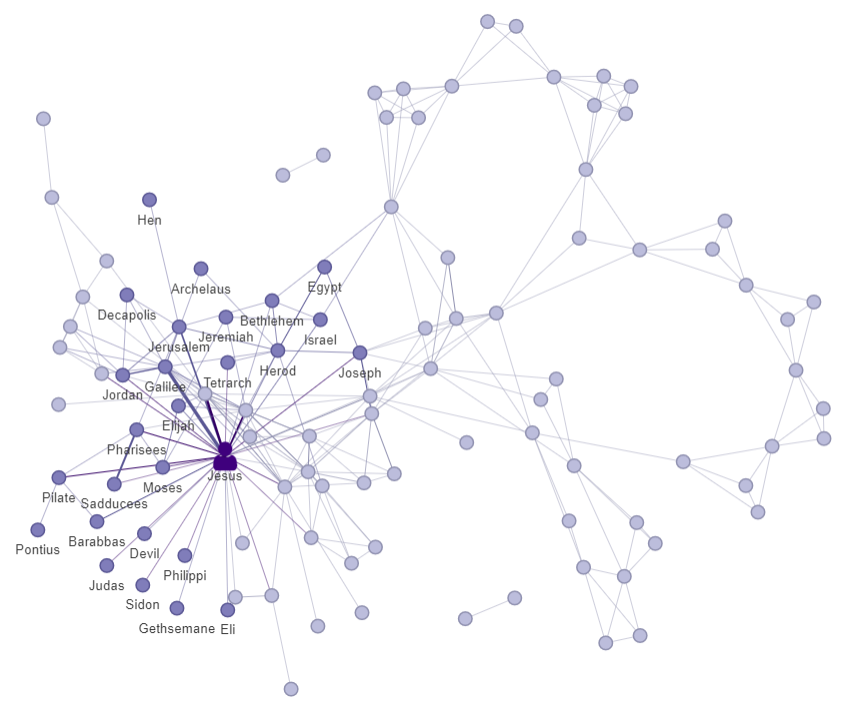}
    \subcaption{\small{\textit{Gospel of Matthew}}}
    \label{networks1}
  \end{subfigure}
  \begin{subfigure}[b]{0.5\textwidth}
    \includegraphics[width=\linewidth]{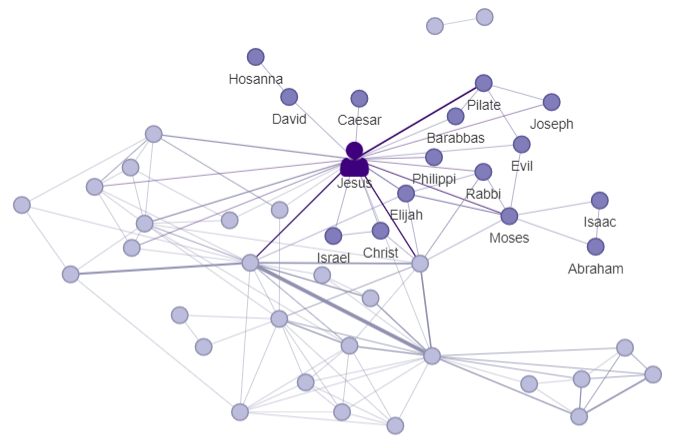}
    \subcaption{\small{\textit{Gospel of Mark}}}
    \label{networks2}
  \end{subfigure}
  

\bigskip
\bigskip 
  
\caption{\small{\textit{The graphs contain the networks in which are depicted the links between names (people and places) in the Acts of Apostles and each Gospel. The highlighted nodes and connections are the communities to which Jesus belongs.}}}

\end{figure}
  
\begin{figure}[ht]\ContinuedFloat
  
  \begin{subfigure}[b]{0.5\textwidth}
    \includegraphics[width=\linewidth]{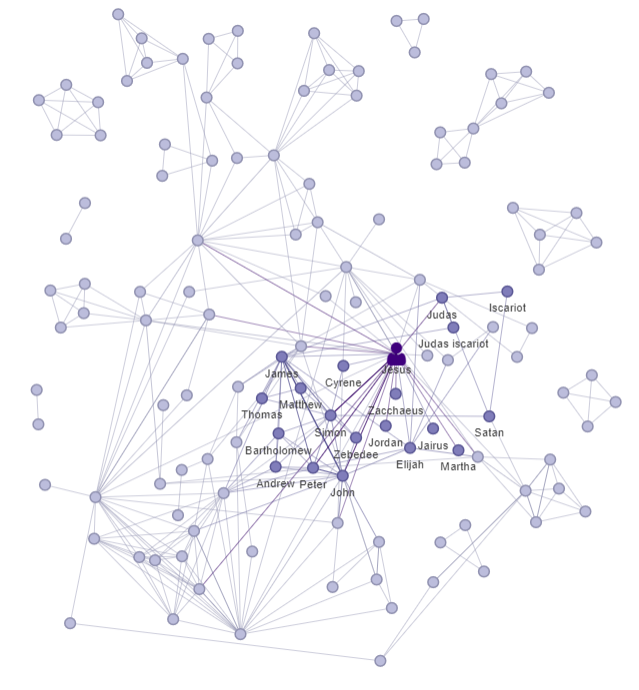}
    \subcaption{\small{\textit{Gospel of Luke}}}
    \label{networks3}
  \end{subfigure}
  \begin{subfigure}[b]{0.5\textwidth}
     \includegraphics[width=\linewidth]{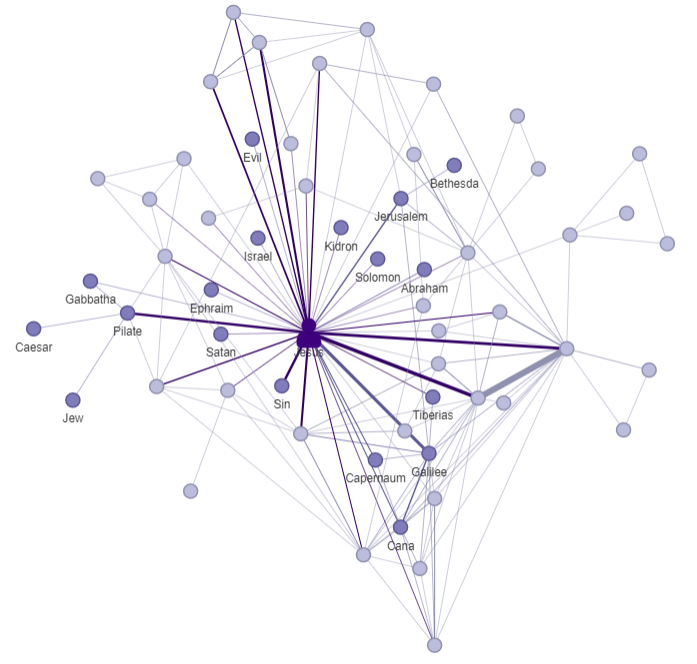}
     \subcaption{\small{\textit{Gospel of John}}}
     \label{networks4}
  \end{subfigure}

\bigskip

\centering

  \begin{subfigure}{0.65\textwidth}
    \includegraphics[width=\linewidth]{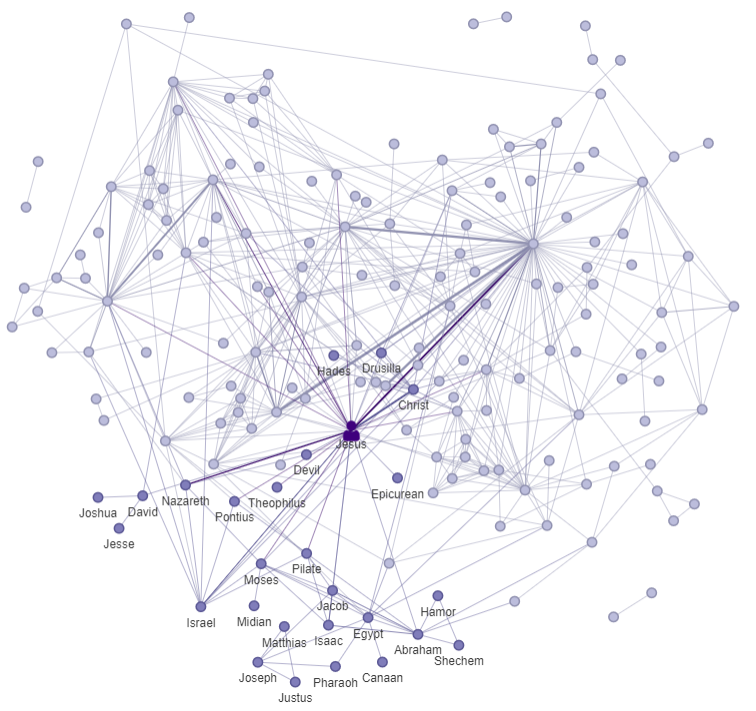}
    \subcaption{\small{\textit{Acts of Apostles}}}
    \label{networks5}
  \end{subfigure}

\bigskip

\caption{\small{\textit{The graphs contain the networks in which are depicted the links between names (people and places) in the Acts of Apostles and each Gospel. The highlighted nodes and connections are the communities to which Jesus belongs. (cont.)}}}
\label{networks}

\end{figure}

\clearpage

\renewcommand{\refname}{

\end{document}